\documentclass[preprint,nofootinbib,amsmath,amssymb,aps,prd,superscriptaddress,a4paper]{revtex4-1}
\pdfoutput=1

\usepackage{graphicx}
\usepackage{dcolumn}
\usepackage{bm}
\usepackage{slashed}
\usepackage{color}
\usepackage{physics}
\usepackage{comment}
\usepackage{cancel}
\usepackage[normalem]{ulem}

\begin{document}
\title{Leptogenesis from magnetic helicity of gauged $\text{U}(1)_{B-L}$}
\author{Hajime Fukuda}
\email{hfukuda@hep-th.phys.s.u-tokyo.ac.jp}
\affiliation{Department of Physics, The University of Tokyo, Tokyo 113-0033, Japan}
\author{Kohei Kamada}
\email{kohei.kamada@ucas.ac.cn}
\affiliation{School of Fundamental Physics and Mathematical Sciences, Hangzhou Institute for Advanced Study, University of Chinese Academy of Sciences (HIAS-UCAS), Hangzhou 310024, China}
\affiliation{International Centre for Theoretical Physics Asia-Pacific (ICTP-AP), Beijing/Hangzhou, China}
\affiliation{Research Center for the Early Universe (RESCEU), Graduate School of Science, The University of Tokyo, Hongo 7-3-1 Bunkyo-ku, Tokyo 113-0033, Japan}
\author{Thanaporn Sichanugrist}
\email{thanaporn@hep-th.phys.s.u-tokyo.ac.jp}
\affiliation{Department of Physics, The University of Tokyo, Tokyo 113-0033, Japan}

\date{\today}
\begin{abstract}
If the $B-L$ symmetry is gauged with the addition of right-handed neutrinos, the standard model $B-L$ current is anomalous with respect to the $B-L$ gauge field itself. Then, the anomaly relation implies that the magnetic helicity of the $B-L$ gauge field is related to the standard model $B-L$ charges, although the whole universe is $B-L$ neutral with right-handed neutrinos.
Based on this, we propose a new leptogenesis scenario with the gauged $B-L$ symmetry as follows. First, the magnetic helicity of the $B-L$ gauge field is generated, e.g., by the axion inflation, together with the standard model and right-handed neutrino $B-L$ charges, with the net $B-L$ charge kept zero. The $B-L$ charges in the standard model and right-handed neutrino sectors are then subject to washout effects from the interactions between them. After the washout effects decouple, the $B-L$ gauge symmetry is Higgsed and the magnetic helicity of the $B-L$ gauge field decays and generates 
$B-L$ charges in the both sector;
thanks to the washout effects, we obtain a non-zero $B-L$ asymmetry.
We show that the baryon asymmetry of the universe can be generated in this scenario, discussing the decay of the magnetic helicity of the $B-L$ gauge field and the interactions between the right-handed neutrinos and the standard model particles.
\end{abstract}

\maketitle

\newcommand{\rem}[1]{{$\spadesuit$\bf #1$\spadesuit$}}
\renewcommand{\theequation}{\thesection.\arabic{equation}}

\renewcommand{\thefootnote}{\fnsymbol{footnote}}
\setcounter{footnote}{0}

\def\thefootnote{\fnsymbol{footnote}}

\section{Introduction}
Understanding the origin of the baryon asymmetry of the universe\,\cite{Planck:2018vyg} is a fundamental challenge in cosmology and particle physics. Various mechanisms have been proposed, including GUT baryogenesis\,\cite{Yoshimura:1978ex,Dimopoulos:1978kv,Toussaint:1978br,Weinberg:1979bt,Barr:1979ye}, thermal leptogenesis\,\cite{Fukugita:1986hr}, and electroweak baryogenesis\,\cite{Kuzmin:1985mm,Arnold:1987mh,Farrar:1993sp}.
Recently, increasing efforts and interests have been devoted to the baryogenesis mechanisms via the decay of magnetic helicity of the $\text{U}(1)_Y$ gauge field in the standard model\,\cite{Giovannini:1997gp,Giovannini:1997eg,Kamada:2016eeb,Kamada:2016cnb}.
By virtue of the chiral anomaly of the $B+L$ symmetry, the change of the magnetic helicity of the $\text{U}(1)_Y$ gauge field corresponds to the change of the $B+L$ number of the universe. 
The magnetic helicity of the $\text{U}(1)_Y$ gauge field inevitably changes during the electroweak phase transition.
Consequently, baryon asymmetry can be generated at that time.
However, in the original scenario, the precise amount of the $B+L$ number depends on the details of the dynamics of the electroweak phase transition; the $B+L$ symmetry is also anomalous with respect to the $\text{SU}(2)_L$ gauge group, and the sphaleron process washes out the $B+L$ number\,\cite{Kuzmin:1985mm}. The final baryon asymmetry is determined by the balance between the generation of the $B+L$ number from the magnetic helicity decay and the washout effects from the sphaleron process.  Thus, the final baryon asymmetry is not solely determined by the magnetic helicity alone.

$B-L$ charges are not washed out by the standard model process as the $B-L$ symmetry in the standard model is not anomalous with respect to $\text{SU}(3)_C\times\text{SU}(2)_L\times\text{U}(1)_Y$. This, however, seems to conclude that $B-L$ cannot be generated with the magnetic helicity, at first glance.
Ref.\,\cite{Domcke:2020quw,Domcke:2022kfs} has circumvent the apparent contradiction by introducing the right-handed neutrinos, which explains the neutrino mass\,\cite{ParticleDataGroup:2022pth} through the seesaw mechanism\,\cite{Minkowski:1977sc,Yanagida:1979as,Gell-Mann:1979vob}. The focus of the work is on the conservation of fermion chiralities in the standard model; some of the standard model Yukawa couplings, such as those for $e, u$, and $d$, are small enough at high temperatures and their chiralities are effectively conserved in the early universe.
The fermion chiralities are anomalous with respect to the standard model gauge groups and can be generated associated with the magnetic helicity.
$B-L$ violating interactions of the right-handed neutrinos, 
which become effective only when they are sufficiently massive before decay, can generate a non-zero $B-L$ in the thermal bath 
with non-zero fermion chirality,
even when the Sakharov conditions\,\cite{Sakharov:1967dj} are not simultaneously satisfied. Note that, however, this scenario depends on the flavor and chirality structure in the UV physics; many extensions of the standard model predict different flavor and chirality structures\,\cite{Isidori:2010kg}.

In this work, we present a new avenue for the generation of the $B-L$ asymmetry from magnetic helicity. Previous works assume that the $B-L$ symmetry is not broken by the anomaly of the standard model gauge groups. However, this is not precise; the $B-L$ symmetry has the gravitational anomaly and the 't Hooft anomaly\,\cite{Weinberg:1996kr}.
The former is considered in the gravi-leptogenesis scenario\,\cite{Alexander:2004us}, but, as Ref.\,\cite{Alexander:2004us} itself admits, the amount of the gravitational wave would be too much if we assumed all the baryon asymmetry of the universe is generated with helical gravitational waves. See also Refs.~\cite{Fischler:2007tj,Kamada:2020jaf}.

On the other hand, the latter possibility has not been explored much, to our knowledge\footnote{See Ref.~\cite{Chao:2024fip} for a recent study to generate baryon asymmetry through the $B-L$ 't Hooft anomaly in axion inflation.}. That is, once the $B-L$ is gauged with the addition of right-handed neutrinos, the standard model $B-L$ symmetry is anomalous with respect to the gauge field itself.  
Based on this, we propose a new mechanism for the generation of the baryon asymmetry of the universe as follows. The magnetic helicity of the $B-L$ gauge field is generated in the early universe, e.g., during the axion inflation, and the $B-L$ charges in the standard model sector are generated with the magnetic helicity. 
The $B-L$ charges in the right-handed neutrino sector 
are also generated with the magnetic helicity, and the total amount of the $B-L$ charges in the universe is zero. By the interactions between the right-handed neutrinos and the standard model particles, the $B-L$ charges in both sectors once tend to be equilibrated and relaxed to zero while keeping the magnetic helicity of the $B-L$ gauge field conserved. Then, as the $B-L$ gauge symmetry is broken at the scale $v_{B-L}$, the magnetic helicity of the $B-L$ gauge field decays and the difference between the $B-L$ charges in the standard model sector and right-handed neutrino sector arises once more. 
If the washout interaction for the $B-L$ charge in the standard model sector
is decoupled before the decay of the magnetic helicity, the difference remains as the baryon number of the universe. As our scenario directly generates $B-L$ charges, we may generate enough baryon asymmetry even if the flavor symmetry is badly broken in the UV physics.

The organization of the paper is following.
In Sec.\,\ref{sec:helicity}, we review the co-generation of the magnetic helicity and chirality and discuss the decay of the former.
In Sec.\,\ref{sec:majorana}, we assume the interaction between the right-handed neutrinos and the standard model particles is such that the standard model neutrinos are Majorana particles and discuss the evolution of the $B-L$ charges in the standard model sector. We find that the result depends on the UV structure of the theory and the phase transition of the $B-L$ gauge symmetry breaking.
In Sec.\,\ref{sec:dirac}, we assume the standard model neutrinos are Dirac particles and discuss the evolution of the $B-L$ charges in the standard model sector. We find that the baryon asymmetry of the universe can be easily generated in this case.
Finally, Sec.\,\ref{sec:conc}, is devoted to the discussion and conclusion.

\section{Generation of the helicity and its decay}
\label{sec:helicity}
In this section, we review the generation of the helicity of the gauge field and discuss its decay. The magnetic helicity of the gauge field is defined as
\begin{align}
    \mathcal{H}_{A} \equiv \int d^3x \vec{A} \cdot \vec{B},
\end{align}
where $\vec{A}$ is the spatial component of the gauge field and $\vec{B}$ is the magnetic field. The magnetic helicity is a gauge invariant quantity if the magnetic field dumps off at the spatial infinity. In the following, we assume the $B-L$ symmetry is gauged and consider the magnetic helicity of the $B-L$ gauge field, $X_\mu$, namely, $\mathcal{H}_X$.

The magnetic helicity has a close relation to the chiral anomaly of the gauge field; the change of the magnetic helicity corresponds to the change of the chiral charge of the fermions. If we ignore the neutrino mass, the standard model $B-L$ current, $J^\mu_{B-L, \text{SM}}$, is anomalous with respect to the $B-L$ gauge field\footnote{Here, we take the covariant form of the anomaly\,\cite{Bardeen:1984pm}.};
\begin{align}
    \label{eq:anomaly}
    \partial_\mu J^\mu_{B-L, \text{SM}} = -\frac{3g_{B-L}^2}{16\pi^2} X_{\mu\nu}\tilde{X}^{\mu\nu},
\end{align}
where $g_{B-L}$ is the coupling constant of the $B-L$ gauge field, $X_{\mu\nu}$ is the field strength tensor of $X_\mu$, and $\tilde{X}^{\mu\nu}$ is the dual tensor of $X_{\mu\nu}$. Integrating the above equation over the spatial volume, we obtain
\begin{align}
    \label{eq:csconservation}
    \frac{d}{dt}\qty(N_{B-L, \text{SM}} + \frac{3g_{B-L}^2}{8\pi^2}\mathcal{H}_{X}) = 0,
\end{align}
where $N_{B-L, \text{SM}}$ is the $B-L$ charge of the standard model particles in the volume. Here, we assume that all fields drop off quickly enough at the spatial infinity.
This equation shows that the change of the magnetic helicity of the $B-L$ gauge field corresponds to the change of the $B-L$ charge of the standard model particles.

In the following, we focus mainly on the dynamics of the $B-L$ gauge field 
in the context of axion inflation. In the subsequent sections, we will analyze the 
evolution of the $B-L$ asymmetry in the matter sector, taking into account
the generation of $B-L$ helicity in axion inflation.  
For this purpose, we need to specify the model of the right-handed neutrinos. 

\subsection{Generation of the helicity}
Let us now review the generation of the helicity of the gauge field. 
The simplest way to generate the helicity is to consider the axion inflation\,\cite{Turner:1987bw,Garretson:1992vt,Anber:2006xt,Barnaby:2010vf,Barnaby:2011vw}. Here, we review it following the ideas given in Refs.~\cite{Jimenez:2017cdr,Domcke:2018eki}. See also Ref.~\cite{Kamada:2022nyt} for a review. 

The Lagrangian of the inflaton $\phi$ and the $B-L$ gauge field is given by
\begin{align}
    \frac{\mathcal{L}}{\sqrt{-g}}= \frac{1}{2}  \partial_\mu\phi\partial^\mu\phi- V(\phi)  -\frac{1}{4}  X_{\mu\nu}X^{\mu\nu}+ \frac{g_{\phi XX}}{4} \phi X_{\mu\nu}\tilde{X}^{\mu\nu} 
\end{align}
where $g_{\phi XX}$ is coupling constant, and we adopt the conformal metric, $ds^2=a^2(t) (d\eta^2-d\vec{x}^2)$ where $\eta$ is conformal time, $a(t)$ is scale factor, and the index, e.g., $\mu$ runs for $\eta,x,y,z$. 
Here the field strength tensor and its dual are defined covariantly, namely,
$X^{\mu\nu}=g^{\mu\rho}g^{\nu\sigma} X_{\rho \sigma}$ and $\tilde{X}^{\mu\nu}=\frac{1}{2\sqrt{-g}}\epsilon^{\mu\nu\rho\sigma}X_{\rho \sigma}$
with $\epsilon^{\mu\nu\rho\sigma}$ being the totally asymmetric Levi-Civita symbol, $\epsilon^{0123}=+1$.
We assume the inflation potential $V(\phi)$ such that $\dot{\phi}$ is almost constant and its change is much slower than the change of the gauge field; this assumption can be satisfied in slow-roll inflation.
We discuss the gauge field $X_\mu=(X_0,\vec{X})$ with the radiation gauge $X_0=0, \ \nabla\cdot \vec{X}=0$ and  perform mode expansion of in the circular polarization basis
as
\begin{equation}
    \vec{X}(\eta, \vec{x})=\int \frac{d^3 k}{ (2\pi)^{3/2}} \sum_{\sigma=\pm} \left[ \vec{\epsilon}^{\, (\sigma)} (\vec{k}) a_{\vec{k}}^{(\sigma)} X_{\sigma} (\eta,\vec{k})  e^{i\vec{k}\cdot \vec{x}}  + \mathrm{h.c.} \right],
\end{equation}
where polarization vector $\vec{\epsilon}^{\, (\pm)}$ satisfying $\vec{\epsilon}^{\, (\pm)}(\vec{k}) \cdot \vec{k}=0,\ i \vec{k} \times \vec{\epsilon}^{\,(\pm)} (\vec{k}) =\pm  k \vec{\epsilon}^{\, (\pm)}(\vec{k})
$ and $\vec{\epsilon}^{\,  (\sigma)*} (\vec{k}) \cdot\vec{\epsilon}^{\, (\sigma')} (\vec{k})= \delta^{\sigma \sigma'}$ with $k=|\vec{k}|$. 
Regarding quantization, we require  
the creation and annihilation operators, $a_{\vec{k}}^{(\sigma)}$  and $a_{\vec{k}}^{(\sigma) \dagger}$, to satisfy the usual canonical commutation relations,
$[a^{(\sigma)}_{\vec{k}},a_{\vec{q}}^{(\sigma') \dagger}]= \delta^{\sigma \sigma'} \delta(\vec{k}-\vec{q})$.

While in the realistic case we need to take into account the induced current from non-perturbative fermion production, let us first neglect it to investigate the dynamics of the gauge field amplification.
The equation of motion for the mode function in the inflationary background then reads 
\begin{equation}
    0=[\partial_{\eta}^2 + k(k \mp 2  \xi  a H_{\rm inf }) ]X_{\pm}(\eta,k), \label{modeeq}
\end{equation}
where 
\begin{equation}
    \xi\equiv - \frac{g_{\phi XX}   \dot{\phi}}{2 H_{\rm inf }}
\end{equation}
is the instability parameter and  $H_{\rm inf}$ is the Hubble constant during inflation. 
We can easily see
that $\pm$ polarization has instability for $\xi \gtrless 0$ and grows exponentially at $k/a < 2 |\xi| H_\mathrm{inf}$. Since $H_{\rm inf }$ and $\xi$ vary only slowly with time during the slow-roll inflation, let us take them constants, 
such that
$a(t)=e^{H_{\rm inf } t}$ and $\eta=-1/aH_{\rm inf }$. 
We find the analytical solution of Eq.~\eqref{modeeq} in this approximation as~\cite{Jimenez:2017cdr,Domcke:2018eki} 
\begin{equation}
    X_{\sigma}(\eta,\vec{k})= \frac{e^{\sigma \pi \xi/2}}{\sqrt{2k}} W_{- i\sigma \xi, 1/2 }(2 ik \eta)
\end{equation}
where $W_{\kappa,\mu}$ is the Whittaker function with $\sigma=\pm$,  
and we have taken the Bunch-Davis vacuum, $ \lim_{-k\eta \rightarrow \infty}  X_{\sigma}(\eta,\vec{k})= e^{-ik\eta}/\sqrt{2k}$.
The exponentially amplified mode is $\pm$ polarization for $\xi \gtrless 0$.
The \emph{physical} electric and magnetic fields are given by
$
    \vec{E}_X = - \partial_\eta \vec{X}/a^2 , \ \vec{B}_X=  \vec{\nabla}\times \vec{X}/a^2.
$
From now on, we assume that $\xi>0$ which is going to ensure the sign of the resultant baryon asymmetry to be positive and correct, as we will see.
Substituting the solution into the mode expansion, for $\xi>0$,  we obtain the physical electromagnetic field as
\begin{gather}
    \langle \vec{E}_X^2 \rangle =\frac{1}{2a^4} \int \frac{d^3 \vec{k}}{(2\pi)^3} \ |\partial_\eta X_{+}|^2\simeq 2.6 \times 10^{-4} \frac{e^{2\pi \xi}}{\xi^3 }H_{\rm inf }^4, \label{eq:axion_EX}\\
    \langle \vec{B}_X^2\rangle =\frac{1}{2a^4} \int \frac{d^3 \vec{k}}{(2\pi)^3} k^2 \ | X_{+}|^2\simeq 3.0 \times 10^{-4} \frac{e^{2\pi \xi}}{\xi^5 }H_{\rm inf }^4 \label{eq:axion_BX}\\
    \langle \vec{E}_X\cdot \vec{B}_X \rangle = -\frac{1}{2a^4}\int \frac{d^3k}{(2\pi)^3} \ k \partial_\eta |X_{+}|^2 \simeq - 2.6 \times 10^{-4} \frac{e^{2\pi \xi}}{\xi^4 }H_{\rm inf }^4 ,
\end{gather}
for sufficiently large $\xi$ {($ \gtrsim 4$~\cite{Jimenez:2017cdr})},
where $\langle\cdot\rangle$ denotes the quantum vacuum expectation value and the cutoff of the momentum 
is taken to be $2\xi a H_\mathrm{inf}$.  See also Ref.~\cite{Ballardini:2019rqh} for the renormalization of the gauge fields in axion inflation. We emphasize that the dominant contribution comes from superhorizon mode $k\lesssim a H_{\rm inf } $, so that the produced gauge field can be regarded to be constant over the Hubble scale, while the electric field and magnetic field point in an anti-parallel direction.
The production rate of the physical helicity is given by
\begin{equation}
     \partial_\eta (a^3 \langle \mathcal{H}_X \rangle) = -2 \int d^3 x a^4 \langle \vec{E}_X\cdot \vec{B}_X\rangle, 
\end{equation}
with which we can identify that a positive magnetic helicity is induced at the end of the inflation.
The spatial average of the magnetic helicity or the magnetic helicity density, $h_X \equiv \langle {\cal H}_X \rangle/\mathbb{V}$ with $\mathbb{V} = a^3\int d^3 x  $ being the volume of spatial hypersurface in Friedmann– Lema\^itre-Robertson-Walker coordinates, at the end of inflation is roughly estimated as
\begin{equation}
    h_X \sim - \left.\frac{2 \langle \vec{E}_X\cdot \vec{B}_X \rangle }{3H}\right|_{\mathrm{end}} \simeq 1.7 \times 10^{-4} \frac{e^{2 \pi \xi}}{\xi^4} H_\mathrm{inf}^3. \label{helinfend}
\end{equation}

Next, we discuss the effect when there is, e.g, a Dirac fermion $\psi$ interacting with gauge field during the inflation through $-g_{B-L} q_\psi \bar{\psi} \gamma^\mu X_\mu\psi$ where $q_\psi$ is the $B-L$ charge of this fermion. 
The amplified gauge fields induce the current of fermions, which backreacts on the dynamics of gauge fields.
The extension to include many fermion species can be done by summing up their contributions. With one Dirac fermion species, the equation governing the energy transfer of the gauge field is given by
\begin{equation}
    \dot{\rho}_X=-4 H_{\mathrm{inf}} \rho_X - 2 \xi H_{\rm inf}  \langle \vec{E}_X\cdot \vec{B}_X \rangle -  g_{B-L} q_\psi \langle \vec{E}_X\cdot \vec{J}_\psi\rangle, \label{eq:EOMrhoX}
\end{equation}
which is derived from {the equation of motion with the fermion current}. 
Here, $\rho_X$ is the expectation value of the energy density of the gauge field, $\rho_X= (1/2 \mathbb{V}) \int d^3 x \langle \vec{E}_X^2  + \vec{B}_X^2 \rangle $, and $\vec{J}_\psi$ is the induced current of fermion $\psi$. 
While the gauge fields evolve as stochastic variables~\cite{Fujita:2022fit}, 
they can be approximated as constant and anti-parallel electric and magnetic fields at the horizon scale, as previously discussed. 
In the presence of a constant magnetic field background, the spectrum of fermions forms Landau levels, with the lowest level exhibiting chiral asymmetry and the higher level exhibiting chiral symmetry.
When electric fields are applied in parallel to magnetic fields,
states are excited 
along the Landau levels, inducing a fermion current~\cite{Nielsen:1983rb,Fukushima:2008xe}.
The induced current 
arises from contributions of the lowest Landau level,  
which describes the chiral anomaly~\cite{Nielsen:1983rb}, 
and from the higher Landau levels, which account for the Schwinger effect~\cite{Heisenberg:1936nmg,Schwinger:1951nm}, representing tunneling processes between the
discrete energy levels. 
Collectively, by taking $\vec{E}_X=(0,0,E_X)$ and $\vec{B}_X=(0,0,-B_X)$ this current is evaluated as~\cite{Domcke:2018eki}
\begin{equation}
  g_{B-L}q_\psi \langle  J^z_\psi \rangle = \frac{(g_{B-L} q_\psi )^3}{6 \pi^2} \coth \left(\frac{\pi B_X}{E_X} \right) E_X B_X \frac{1}{H_{\rm inf }}, \label{eq:Jpsi}
\end{equation}
where scattering among particles is neglected.
In the case 
of the standard model contents with, e.g., the addition of
3 generation right-handed neutrinos, we can modify Eq.~\eqref{eq:Jpsi} as
\begin{equation}
    q^3_\psi \rightarrow  3 \left( 3\times  \left(\frac{1}{3}\right)^3+3\times \left(\frac{1}{3}\right)^3+1+1\right),
\end{equation}
where the first factor of 3 outside big parenthesis accounts for 3 generations. The first and the second contributions come from up and down quarks of each color, while the remaining terms are from electrons and neutrinos. Here, for simplicity, we assume the $B-L$ charge of the right-handed neutrinos is $1$.

Let us now examine how to include the effect of the induced current. 
While there are several approaches on this problem~\cite{Domcke:2018eki,Gorbar:2021rlt,Gorbar:2021zlr,Fujita:2022fwc}, 
no quantitatively precise consensus has been obtained. 
Here we adopt the ``equilibrium estimate''~\cite{Domcke:2018eki} as an example to 
give a concrete estimate.
Its idea is described as follows. 
In case there is no fermion, the energy feeding from the axion is balanced by the cosmic expansion giving a constant electromagnetic field configuration.
On the other hand, in the presence of 
charged
fermions, 
the induced current leads to additional
energy transfer from the gauge field to the fermion sector, 
as reflected in
the last term of the right-hand side of Eq.~\eqref{eq:EOMrhoX}. Then, we expect the dynamical equilibrium $\dot{\rho}_X=0$, meaning that the energy in gauge field sector is
balanced between the energy feeding from axion dynamics
and the energy drain to  both the fermion sector and cosmic expansion, 
\begin{equation}
    0=-2 H_{\rm inf } \langle \vec{E}_X^2+\vec{B}_X^2 \rangle -2 \xi H_{\rm inf}  \langle \vec{E}_X\cdot \vec{B}_X \rangle -  g_{B-L} q_\psi \langle \vec{E}_X\cdot \vec{J}_\psi\rangle,
    \label{eq:backreaction_equilibrium}
\end{equation}
Approximating $\langle \vec{E}_X^2 \rangle \simeq E_X^2, \langle \vec{B}_X^2 \rangle \simeq B_X^2,  \langle \vec{E}_X\cdot \vec{B}_X \rangle \simeq -E_X B_X,$ and $ \langle \vec{E}_X\cdot \vec{J}_\psi\rangle \simeq E_X \langle  J^z_\psi \rangle$, and substituting induced current, Eq.~\eqref{eq:Jpsi}, into Eq.~\eqref{eq:backreaction_equilibrium}, we obtain 
\begin{equation}
    0=-2 H_{\rm inf } (E_X^2+B_X^2)+2\xi_{\rm eff} H_{\rm inf } E_X B_X, \label{eq:equilibriumxieff}
\end{equation}
with
\begin{equation}
    \xi_{\rm eff} \equiv \xi-\frac{(g_{B-L} q_\psi )^3}{12 \pi^2} \coth \left(\frac{\pi B_X}{E_X} \right) \frac{E_X }{ H_{\rm inf }^2} \label{eq:xieff}
\end{equation}
denoting the effective instability parameter. 
The effect of fermion production
is now implemented
as a suppressed production of the helicity of the gauge field.
We then estimate the resultant gauge field strength as follows.
By assuming constant electromagnetic configuration,  $\xi_{\rm eff}$ is also taken to be constant. 
{Let us assume that $\xi_{\rm eff}$ would be an independent and constant parameter. 
The equation of motion of $\vec{X}$ with the fermion current is identical to Eq.\,\eqref{modeeq}, where we do not have any fermion, with the replacement of $\xi\rightarrow \xi_{\rm eff}$.
Therefore, the solutions for $E_X$ and $B_X$ would take the same form as the analytical solutions in  Eqs.~\eqref{eq:axion_EX} and \eqref{eq:axion_BX}, with $\xi$ replaced by $\xi_{\rm eff}$. In reality, $\xi_{\rm eff}$ is a dependent and dynamical parameter, but once the system is in equilibrium and $\xi_{\rm eff}$ becomes constant, the values of $E_X$ and $B_X$ are expected to approach those values. With that solution for $E_X$ and $B_X$, the second term of the right-hand side of Eq.\,\eqref{eq:xieff} is now a function of $\xi_{\rm eff}$ and Eq.\,\eqref{eq:xieff} can be now regarded as a self-consistency equation on $\xi_{\rm eff}$ in terms of $\xi$. Thereby, we may estimate $\xi_{\rm eff}$, $E_X$ and $B_X$. This is called the ``equilibrium estimate''.
One may worry about the strong backreaction on the inflaton dynamics~\cite{Garcia-Bellido:2023ser,Figueroa:2023oxc,vonEckardstein:2023gwk,Figueroa:2024rkr}, but if $\xi_\mathrm{eff}$ is small, regardless of the value of $\xi$, the backreaction is negligible and our treatment is consistent. }  
The magnetic helicity at the end of inflation is also evaluated by Eq.~\eqref{helinfend} with $\xi$ replaced by $\xi_\mathrm{eff}$.
{Note that, there could be other constant solutions if $E_X$ and $B_X$ do not take the same form as Eqs.~\eqref{eq:axion_EX} and \eqref{eq:axion_BX}. In that case, the magnetic helicity becomes larger than the equilibrium estimate\,\cite{Gorbar:2021zlr} and the equilibrium estimate, which we adopt in this paper, is expected to be more conservative. \footnote{{All the estimates for the Schwinger backreaction thus far rely on the approximation that decomposes the electric and magnetic field in the expression of Schwinger current into the background and dynamical field, which is not very correct. To obtain a more precise estimate, numerical simulations with more appropriate treatment of the Schwinger current are needed.}}}
As has been discussed, the produced fermions carry $B-L$ asymmetries. 
From Eq.~\eqref{eq:csconservation}, we evaluate the $B-L$ number
density in the standard model sector at the end of inflation as
\begin{equation}
    n_{B-L,SM} \equiv \frac{N_{B-L,SM}}{\mathbb{V}}= - \frac{3 g_{B-L}^2}{8 \pi^2} h_X \simeq - 0.6 \times 10^{-4} \frac{g_{B-L}^2e^{2 \pi \xi_\mathrm{eff}}}{\pi^{2} \xi^4_\mathrm{eff}} H_\mathrm{inf}^3. 
\end{equation}

\subsection{Evolution of the helicity}
Next, we discuss the dynamics of the plasma with $B-L$ interaction and the $B-L$ gauge field
after inflation, with instant reheating in mind. In this subsection, we focus on the case where the $B-L$ gauge field is massless.
Throughout this paper, we require that the magnetohydrodynamics (MHD) description is valid for the $B-L$ gauge field and the magnetic helicity is conserved if the $B-L$ symmetry is not broken. We clarify the necessary conditions for these requirements in the following discussion.
As is the case for the standard model hypercharge gauge field\,\cite{Giovannini:1997eg,Son:1998my,Jedamzik:1996wp,Banerjee:2004df,Domcke:2019mnd}, the following MHD equation is to be satisfied;
\begin{align}
    \vec{J}_{B-L} &= \sigma (\vec{E}_X + \vec{v}\times\vec{B}_X), \label{eq:ohm}\\
    \nabla\times\vec{B}_X &= \vec{J}_{B-L} \label{eq:Ampere},\\
    \frac{\partial}{\partial t} \vec{v} + (\vec{v}\cdot\nabla)\vec{v} &= \frac{\eta_\mathrm{vis}}{\rho + p} \nabla^2 \vec{v} + \frac{1}{\rho + p}\qty(\vec{J}_{B-L}\times\vec{B}_X), \label{eq:navierstokes}
\end{align}
where $\vec{J}_{B-L}$ is the total $B-L$ current, $\sigma$ is the $B-L$ conductivity, $\vec{v}$ is the velocity of the fluid, $\eta_\text{vis}$ is the shear viscosity, $\rho$ is the energy density of the fluid, and $p$ is the pressure of the fluid. We have ignored the chiral magnetic effect~\cite{Fukushima:2008xe,Son:2009tf,Neiman:2010zi} which does not change 
the evolution of the system unless the chiral plasma instability~\cite{Joyce:1997uy,Akamatsu:2013pjd,Rogachevskii:2017uyc,Schober:2017cdw,Kamada:2018tcs} or the anomalous chirality cancellation~\cite{Brandenburg:2023rul,Brandenburg:2023aco} becomes effective. In order to explain the present baryon asymmetry of the universe, the mechanism discussed in the subsequent subsections should take place.
The $B-L$ magnetic field must be subdominant to the radiation energy density of the plasma, $B_X^2 \ll \rho$. For simplicity, we assume all particles in the plasma are massless and in thermal equilibrium. Additionally, we assume that the typical scale of the dynamics, $L$, is much larger than the mean-free path of the $B-L$ charge carriers. 
The cosmic expansion does not appear in the MHD equations, 
because for massless gauge fields it can be removed by moving
to the conformal frame~\cite{Brandenburg:1996fc}. 
Here we use a conventional treatment 
following Ref.~\cite{Domcke:2019mnd}, 
which assumes that the typical spatial scales for the magnetic and velocity fields are the same. 
Although recent studies show that it is not always the case~\cite{Uchida:2022vue,Uchida:2024ude}, this approach is enough to give a rough estimate.

Let us consider the validity of these equations. The first equation is the Ohm's law for the $B-L$ current and can be qualitatively understood from the Drude model\,\cite{Arnold:2000dr}. In the Drude model, the conductivity of the $B-L$ current in the conformal frame is 
\begin{align}
    \sigma \simeq a(T) \sum_i q_{i}^2 g_{B-L}^2 g_i T^2 \tau_i,
\end{align}
where $a(T)$ is the scale factor of the universe at $T$, $q_{i}$ is the $B-L$ charge of the charge carrier $i$, $T$ is the temperature of the universe,
$g_i$ is the number of the internal degrees of freedom of $i$ and $\tau_i$ is the mean-free time for $i$.
For the standard model particles, the dominant contribution comes from the right-handed charged leptons, where $\tau \sim 1/(g'^4 T)$, where $g'$ is the $\text{U}(1)_Y$ gauge coupling constant.
With new particles, such as a sterile neutrino, which interacts with the standard model particles only via the $B-L$ gauge field, $\tau$ can be longer and $\sigma$ can be bigger. However, as we have assumed $L\gg \tau$, there is an upper bound for $\sigma$, $\sigma \lesssim a(T) g_{B-L}^2 T^2 L.$ Thus, as a typical value, we 
take $\sigma \sim a(T) g_{B-L}^2 T / g'^4$ in the subsequent discussion.

The second equation is the Amp\`ere's law for the $B-L$ current; it is derived from one of the Maxwell's equations, 
\begin{align}
    \nabla\times\vec{B}_X = \vec{J}_{B-L} + \frac{\partial}{\partial t} \vec{E}_X.
\end{align}
Using the Ohm's law, Eq.\,\eqref{eq:ohm}, we can eliminate the displacement current $\frac{\partial}{\partial t} \vec{E}_X$ if
\begin{align}
    \frac{1}{\sigma T'} & \ll 1, \label{eq:assump_ampere1}\\
    |\vec{v}| \cdot \frac{L}{T'} & \ll 1 \label{eq:assump_ampere2},
\end{align}
where $T'$ is the typical timescale of the dynamics. In the expanding universe, if the magnetic field is created during the inflation, $T'$ is identified as the conformal time, $T' \simeq \frac{a(T)}{a^2(T_\text{re})} H(T_\text{re})^{-1}$, where $a(T)$ is the scale factor of the universe, $H(T)$ is the Hubble parameter, and $T_\text{re}$ is the reheating temperature. In such a case, the first condition is satisfied if
\begin{align}
    \frac{g'^4}{g_{B-L}^2} \ll \frac{M_p}{T_\text{re}} \label{eq:require1},
\end{align}
where $M_p$ is the reduced Planck mass. The second condition is satisfied if $|\vec{v}| \ll 1$, which we will see is satisfied in the later discussion.

The third equation is the Navier-Stokes equation for the fluid and can be derived from the conservation of the energy-momentum tensor of the imperfect fluid\,\cite{Weinberg:1972kfs}. We have assumed that the fluid is uniform, $\rho, p = \text{constant}$. The viscosity is written as\,\cite{Weinberg:1971mx,Arnold:2000dr}
\begin{align}
    \eta_{\mathrm{vis}} = \frac{4}{15}\rho\tau.
\end{align}
Therefore, we 
take the kinetic viscosity, $\nu$, defined as $\nu \equiv \eta_{\mathrm{vis}}/(\rho + p)$, to be around
\begin{align}
    \nu \sim g'^{-4} (a(T) T)^{-1}
\end{align} 
with the contributions from the standard model particles being dominant.

Next, we discuss the dynamics of the magnetic field and the fluid. With the Amp\`ere's law, Eq.\,\eqref{eq:Ampere}, the Ohm's law, Eq.\,\eqref{eq:ohm}, and the other Maxwell equations, the equation of motion of the magnetic field is
\begin{align}
    \frac{\partial}{\partial t}\vec{B}_X = \nabla\times\qty(\vec{v}\times\vec{B}_X) + \frac{1}{\sigma}\nabla^2\vec{B}_X.
\end{align}
The first term on the right-hand side is the advection term and the second term is the diffusion term. If the diffusion term is much larger than the advection term, the equation is the diffusion equation and the magnetic field is quickly damped.
To avoid this, we require the advection term to be much larger than the diffusion term:
\begin{align}
    R_m \equiv \sigma L |\vec{v}| \gg 1 \label{eq:assump_magrey},
\end{align}
where we have defined the magnetic Reynolds number $R_m$. With this requirement, 
the magnetic helicity is indeed conserved:
\begin{align}
    \frac{d}{dt}\mathcal{H}_X 
    &\simeq -\frac{2}{T'} R_m^{-1} \qty(\frac{|\vec{v}|}{L/T'}) \mathcal{H}_X \ll -\frac{1}{T'} \mathcal{H}_X,
\end{align}
just as the case for the standard model hypercharge gauge field\,\cite{Giovannini:1997eg,Son:1998my,Jedamzik:1996wp,Banerjee:2004df,Domcke:2019mnd}.

To estimate the magnetic Reynolds number, we need to know the typical velocity of the fluid. Although we need numerical simulations for more precise evaluations, we can roughly estimate the velocity of the fluid from the Navier-Stokes equation, Eq.\,\eqref{eq:navierstokes}\,\cite{Banerjee:2004df,Domcke:2019mnd}. Using the Amp\`ere's law, Eq.\,\eqref{eq:Ampere}, the Navier-Stokes equation is written as
\begin{align}
    \frac{\partial}{\partial t} \vec{v}  = -\qty(\vec{v}\cdot \vec{\nabla})\vec{v} + \nu \nabla^2 \vec{v} + \frac{1}{\rho + p}\qty(-\frac{1}{2} \vec{\nabla} B_X^2 + \qty(\vec{B}_X\cdot\vec{\nabla})\vec{B}_X).
\end{align}
If the first term on the right-hand side is much larger than the second term, namely the kinetic Reynolds number is much larger than unity, 
\begin{align}
    R_e \equiv \frac{|\vec{v}| L}{\nu} \gg 1,
\end{align}
the velocity of the fluid is determined by the balance between the first and the third terms on the right-hand side as
\begin{align}
    v \sim \frac{B_X}{T^2} \label{eq:v_largere},
\end{align}
and the equipartition between the magnetic energy and the kinetic energy is satisfied. On the other hand, if the kinetic Reynolds number is much smaller than unity, the velocity of the fluid is determined by the balance between the second and the third terms on the right-hand side as
\begin{align}
    v \sim \sqrt{R_e}\frac{B_X}{T^2} \sim \frac{L}{\nu}\frac{B_X^2}{T^4}\label{eq:v_smallre}.
\end{align}
In any case, the velocity of the fluid is much smaller than unity as the magnetic field is subdominant to the radiation energy density of the plasma, $B_X^2 \ll \rho$, and the requirement, 
Eq.\,\eqref{eq:assump_ampere2}, is satisfied.

To summarize this subsection, we would like to clarify the requirements 
we have used.
In the above discussion, we have required the large conductivity, Eq.\,\eqref{eq:assump_ampere1}, the small velocity of the fluid, Eq.\,\eqref{eq:assump_ampere2} and the large magnetic Reynolds number, Eq.\,\eqref{eq:assump_magrey}. 
We have already discussed that the first and the second requirements are easily met; the first requirement is satisfied if the reheating temperature is much larger than the Planck mass, Eq.\,\eqref{eq:require1}, and the second requirement is automatically satisfied as the magnetic field is subdominant to the radiation energy density of the plasma. Assuming that the magnetic field is created during the inflation, the third assumption is reduced to
\begin{align}
    \frac{g_{B-L}^2B_{X0} H(T_\text{re})^{-1}}{g'^4T_\text{re}} \times \min\qty(1, \frac{g'^4 B_{X0} H(T_\text{re})^{-1}}{T_\text{re}}) \gg 1, \label{infmagturb}
\end{align}
where $B_{X0}$ is the amplitude of the magnetic field at the reheating and we have used $L \sim (a(T_\mathrm{re}) H(T_\text{re}))^{-1}$.

\begin{figure}[t]
    \centering
    \includegraphics[width=0.5\textwidth]{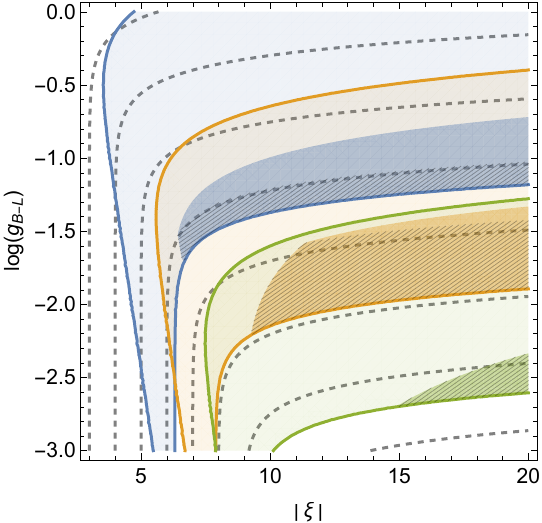}
    \caption{The parameter space where the $B-L$ helicity is well-conserved,
    with the magnetic field created during the inflation. The vertical axis is the effective coupling constant of the $B-L$ gauge field, and the horizontal axis is the instability parameter $\xi$ between the inflaton and the $B-L$ gauge field.
    The black dashed lines are contours of the effective inflaton-$B-L$ instability parameter, $\xi_\text{eff}=3,4,\cdots 10$, from the left to the right. 
    The blue, orange, and green shaded regions correspond to $H_{\mathrm{inf}} = 10^{12}, 10^{10}$ and $10^8\,\text{GeV}$, respectively, where $H_{\mathrm{inf}} $ is the Hubble constant during the inflation. For the details of these regions, see the main text.}
    \label{fig:b-lmhd}
\end{figure}

In Fig.\,\ref{fig:b-lmhd}, we illustrate the parameter space where the
$B-L$ helicity is well-conserved, under the assumption that the magnetic field is generated during inflation and the reheating process is instantaneous. The light-shaded regions indicate where the energy density of the $B-L$ gauge field is sufficiently small (lower boundary) and where the helicity-to-entropy ratio is sufficiently large (upper boundary). Specifically, for the lower boundary, we require that the energy density of the $B-L$ gauge field is less than $1\%$ of the total energy density at the reheating temperature.  For the upper boundary, we define the helicity-to-entropy ratio as
\begin{align}
    \eta_{\mathcal{H}} \equiv \qty|\frac{3 g_{B-L}^2}{8 \pi^2}  \frac{h_X}{s(T_\text{re})}|,
\end{align}
where $s$ is the entropy density of the universe. We require that the helicity-to-entropy ratio is more than $\frac{79}{28} \eta_0$, where $\eta_0 \simeq 8.7 \times 10^{-11}$ is the observed baryon-to-entropy ratio in the current universe\,\cite{Planck:2018vyg}, 
to ensure that the $B-L$ helicity decay would explain 
the present baryon asymmetry of the universe as we will discuss in the next subsection. 
The dark-shaded regions indicate where the 
magnetic Reynolds number is larger than the unity (Eq.~\eqref{infmagturb}) such that the $B-L$ helicity is well-conserved
among the corresponding light-shaded regions. In these regions, we show the region where the kinetic Reynolds number is larger than unity, $R_e > 1$, as the meshed regions.
We also show the contours of the effective inflaton-$B-L$ instability parameter, $\xi_\text{eff} = 3, 4, \cdots, 10$, according to the equilibrium estimation reviewed in the previous subsection.

The contours of $\xi_\text{eff}$ show that the effective instability parameter can be large if the gauge coupling $g_{B-L}$ is small. This is because the backreaction to the effective instability parameter is smaller for smaller $g_{B-L}$. For the larger $\xi_\text{eff}$, the helicity-to-entropy ratio can be larger despite the smaller gauge coupling.
For the $B-L$ helicity
to be well-conserved, 
the magnetic Reynolds number should be much larger than unity, $R_m \gg 1$. This requires that the magnetic field is sufficiently large, and therefore the viable parameter space lies in the smaller $g_{B-L}$ and larger $\xi_\text{eff}$ regions.

Before moving to the next subsection, let us comment on the uncertainty of our discussion about the MHD evolution. 
Since the MHD equations describe a fully non-linear system, 
the estimates are not very precise to determine its evolution. 
While a sufficiently large magnetic Reynolds number is a necessary condition
for the survival of the magnetic helicity, 
our estimate is based on rough order estimates and hence is not quantitatively precise. 
For example, the estimate of the velocity field at a small kinetic Reynolds 
number might not be correct since it is not clear if a turbulent dynamics 
takes place, which is essential for our order estimate. 
This is the reason why Ref.~\cite{Domcke:2022kfs} adopted several ways
to estimate the magnetic Reynolds number. 
Moreover, recent studies suggest that the typical length scale for the 
magnetic field and velocity field might be different~\cite{Uchida:2022vue,Uchida:2024ude}, 
which violates the first assumption of our estimate. 
Therefore, one should regard the parameter space for the survival 
of the $B-L$ helicity in Fig.~\ref{fig:b-lmhd} as just a demonstration
for a rough estimate. 
The origin of this uncertainty comes from the hierarchy between
the magnetic and kinetic Reynolds number, $R_m/R_e \sim g^2_{B-L}/g'^8$, 
for not so small $g_{B-L}$, 
which is difficult to be implemented in the numerical simulation. 
In order to determine precisely the condition for the helicity 
to be well-conserved, more precise studies in the case where 
the magnetic and kinetic Reynolds numbers are hierarchical 
are needed, but it is beyond the scope of the present study. 

\subsection{Decay of the helicity}
Let us discuss the fate of the $B-L$ helicity in the presence of a Higgs field with a negative mass squared.
It is important to note that even if such a Higgs field exists, the phase of the system may differ from the one without helicity due to the anomaly equation, Eq.\,\eqref{eq:csconservation}, which links the helicity to the fermion number\,\cite{Rubakov:1985nk}. However, in our case of interest, where the helicity-to-entropy ratio is much smaller than unity, the phase of the system is not significantly affected by the helicity, and the $B-L$ symmetry is broken.

In the Higgs phase, the magnetic helicity can decay through two mechanisms: the imaginary part of its self-energy and Ohmic dissipation. The former would be 
analogous to the decay of a $B-L$ gauge boson particle in a vacuum. Consequently, we expect the decay rate to be similar to that of the $B-L$ gauge boson in a vacuum. The lifetime $\tau_X$ is thus given by
\begin{align}
    \tau_X^{-1} \simeq \Gamma_X \equiv \frac{g_{B-L}^2\sum q_i^2}{24\pi} m_X.
\end{align}
Here, we assume the fermions are massless.

To estimate the decay rate due to Ohmic dissipation, we first write down the equation of motion for the $B-L$ gauge field and the matter fields, which may replace Eq.~\eqref{eq:ohm} and Eq.~\eqref{eq:Ampere} in the previous subsection.
For the time being, let us consider the Minkowski spacetime, $a = 1$ for simplicity. We will back to this point later.
With the existence of a would-be Nambu-Goldstone boson $\phi_M$ for the $B-L$ gauge symmetry, the effective Lagrangian of the $B-L$ gauge field is
\begin{align}
    \mathcal{L} = -\frac{1}{4}X_{\mu\nu}X^{\mu\nu} + \frac{1}{2}m_X^2 \qty(X_\mu - g_{B-L}^{-1} v_{B-L}^{-1} \partial_\mu \phi_M)^2 - X^\mu J_{B-L, \mu},
\end{align}
where $m_X$ is the mass of the $B-L$ gauge field and $v_{B-L}$ is the VEV of the Higgs field breaking the $B-L$ gauge symmetry\,\footnote{Here, for simplicity, we ignore the Chern-Simons term between $X_\mu$ and $\phi_M$.}. 
The mass of the $B-L$ gauge field is $m_X = q g_{B-L} v_{B-L}$, where $q$ is the $B-L$ charge of the Higgs field.
We may redefine the $B-L$ gauge field as $X_\mu \to X_\mu + g_{B-L}^{-1} v_{B-L}^{-1} \partial_\mu \phi_M$ to get rid of the would-be Nambu-Goldstone boson. The equation of motion of the $B-L$ gauge field is then
\begin{align}
    \nabla \cdot \vec{E}_X &= \rho_{B-L} - m_X^2 X_0, \label{eq:max_pr1}\\
    \nabla \times \vec{B}_X - \frac{\partial}{\partial t} \vec{E}_X &= \vec{J}_{B-L} - m_X^2 \vec{X} \label{eq:max_pr2}.
\end{align}
or, equivalently,
\begin{align}
    \qty(\frac{\partial^2}{\partial t^2} - \nabla^2 + m_X^2)X^\mu = J_{B-L}^\mu\label{eq:max_pr}.
\end{align}
Here, again we have ignored the chiral magnetic effect. In this paper, we assume that the $B-L$ symmetry is Higgsed before the anomalous chirality cancellation~\cite{Brandenburg:2023rul,Brandenburg:2023aco} would occur.
The gauge transformation of the matter fields can absorb the redefinition, and the equations of motion of the matter fields are not changed. Thus, we assume that the Ohm's law, Eq.\,\eqref{eq:ohm}, is still valid after the $B-L$ gauge symmetry is broken. The time dependence of the electric field is not negligible, and the electric conductivity may be different from the one before the $B-L$ gauge symmetry is broken. However, for the time being, we assume $\sigma \sim g_{B-L}^2 T^2 \tau \sim g_{B-L}^2 T / g'^4$ if the temperature $T$ is not much smaller than $v_{B-L}$.
We will return to this point later.

From Eq.\,\eqref{eq:ohm} and Eq.\,\eqref{eq:max_pr}, we can eliminate the current and obtain the time evolution of the gauge field as
\begin{align}
    \qty(\frac{\partial^2}{\partial t^2} - \nabla^2 + m_X^2 + \sigma \frac{\partial}{\partial t}) \vec{X} = -\sigma \vec{\nabla}X^0 + \sigma \vec{v}\times\qty(\vec{\nabla}\times\vec{X}).
\end{align}
To further simplify the equation, we assume the typical spatial scale of the $B-L$ gauge field is much larger than the typical time scale of the $B-L$ gauge field, $L \gg T'$. We will discuss the validity of this assumption later. 
Also, in the situation of our interest, the net $B-L$ charge is zero or, at least, much smaller than the current and we may neglect the term $X^0$.
The equation then becomes
\begin{align}
    \qty(\frac{\partial^2}{\partial t^2} + m_X^2 + \sigma \frac{\partial}{\partial t}) \vec{X} \simeq 0.
\end{align}
This equation is the damped harmonic oscillator equation and the solution is
\begin{align}
    \qty|\vec{X}| \simeq \qty|\vec{X}(0)| e^{-\frac{t}{\tau_{\text{MHD}}}},
\end{align}
where
\begin{align}
    \tau_{\text{MHD}} = \begin{cases}
        \frac{2}{\sigma} \quad &\text{for } 2 m_X > \sigma,\\
        \frac{\sigma + \sqrt{\sigma^2 - 4m_X^2}}{2m_X^2} \quad &\text{for } 2 m_X < \sigma.
        \end{cases}.
\end{align}
Here, for $2 m_X < \sigma$, we have taken a solution with a decreasing lifetime as the mass increases.

Let us now make a few comments.
First, we have assumed that the electric conductivity is not much different from the one before the $B-L$ gauge symmetry is broken. 
However, in the broken phase, the time dependence of the electric field can be much larger than the spatial dependence of the electric field; if the fluid is non-conducting, the time frequency of the electric field is $m_X$. When the gauge field is oscillating, the electric conductivity is different from the one for non-oscillating gauge field. According to the Drude model\,\cite{ashcroft2011solid}, the electric conductivity, $\sigma(\omega)$, for oscillating gauge field with the frequency $\omega$ is now complex and
\begin{align}
    \sigma(\omega) = \frac{\sigma}{1 - i \omega \tau}.
\end{align}
For $T \ll v_{B-L}$, $m_X \tau \gg 1$ and the fluid becomes non-conducting.
Second, we have assumed that the typical spatial scale of the $B-L$ gauge field is much larger than the typical time scale of the $B-L$ gauge field. This is justified for
\begin{align}
    \label{eq:largeEnoughMx}
m_X \gtrsim \sqrt{\frac{|\vec{v}|\sigma}{L}} = \frac{\sqrt{R_m}}{L}.
\end{align}
Otherwise, the decay rate might be less effective.
In the following, however, we assume that the discussion in the previous paragraph is unchanged.

Taking both decay effects of the gauge field itself into account, we, very roughly, conclude the time evolution of the magnetic helicity is given as
\begin{align}
    \mathcal{H}_X \simeq \mathcal{H}_X(0) e^{-\frac{t}{\tau_{\mathcal{H}}}},
\end{align}
where
\begin{align}
    \tau_{\mathcal{H}} \equiv \frac{1}{2}\min\qty(\tau_{\text{MHD}}, \tau_X).
\end{align}
However, we emphasize that for more precise estimation, we need numerical simulations of MHD 
and more careful analyses of the decay rate of the helicity. 

We have so far ignored the cosmic expansion and the $B-L$ phase transition.
Let us estimate when the helicity decays in the cosmic history
and examine its validity. 
Here we assume that the gauge field obtains the mass via the Higgs mechanism,
\begin{align}
    m_X = q g_{B-L} v_{B-L}(T),
\end{align}
where $v_{B-L}(T)$ is the VEV of the Higgs field at the temperature $T$. 
We also assume that the $B-L$ phase transition is of the second order.
$v_{B-L}(T)$ depends on the critical exponent of the $B-L$ phase transition, $\beta$, $v_{B-L}(T) \propto (T_c - T)^\beta$, where $T_c$ is the critical temperature of the $B-L$ phase transition.
Let $T_{HD}$ be the temperature when the helicity decays. As we have discussed in this section, it is difficult to estimate the decay rate of the helicity precisely, but for simplicity, we assume that $T_{HD}$ is the temperature when the helicity decay time is equal to the Hubble time, 
\begin{align}
    \label{eq:decayTime}
    H(T_{HD})^{-1} = \tau_{\mathcal{H}}|_{m_X = q g_{B-L} v_{B-L}}.
\end{align}
Note that near the critical temperature, the universe undergoes a quenching process where the system cannot keep up with the rapid changes in temperature\,\cite{Kibble:1976sj,Zurek:1985qw}. This quenching and the formation of the topological defects could potentially impact the MHD description and the helicity decay. However, for simplicity, we ignore these effects and assume our estimation remains valid.

To evaluate the decay time, let us focus on the decay time of the helicity from the MHD, $\tau_{\text{MHD}}$; we use $H(T_{HD})^{-1} \lesssim \tau_{\text{MHD}} \sim \sigma/m_X^2$. Then, we obtain the following inequality:
\begin{align}
    q \frac{v_{B-L}(T_{HD})}{T_{HD}} \lesssim \frac{1}{g'^2}\sqrt{\frac{H(T_{HD})}{T_{HD}}} \ll 1.
\end{align}
Here, because $L \gtrsim H(T_{HD})^{-1}|\vec{v}|$ due to 
the MHD dynamics~\cite{Banerjee:2004df}\footnote{Here we suppose that the coherence length of magnetic field is larger than the Alfv\'en scale. If we adopt the reconnection-driven turbulence~\cite{Uchida:2022vue,Uchida:2024ude}, the coherence length can be shorter.}, 
the requirement, Eq.\,\eqref{eq:largeEnoughMx}, is satisfied at least for $q \frac{v_{B-L}(T_{HD})}{T_{HD}} \sim \frac{1}{g'^2}\sqrt{\frac{H(T_{HD})}{T_{HD}}}$. 
If we assume the Landau-Ginzburg theory, 
\begin{align}
    v_{B-L}(T) = v_{B-L}^0 \qty|1 - \frac{T}{T_c}|^\frac{1}{2}
\end{align}
with $v_{B-L}^0 \sim T_c$, we can conclude $T_c - T_{HD} \ll T_c$; the helicity quickly decays after the phase transition, in a shorter time scale than the Hubble time. Therefore, we expect that the effect of the cosmic expansion is negligible.
If this condition is not satisfied, for example, if the mass of the $B-L$ gauge field is Stueckelberg-like\,\footnote{The 't Hooft anomaly may be canceled by the four-dimensional analogue of the Green-Schwarz mechanism\,\cite{Green:1984sg,Faddeev:1986pc,Krasnikov:1985bn,Babelon:1986sv,Harada:1986wb,Preskill:1990fr}, but the cutoff scale cannot be arbitrarily large in that case.} and no phase transition occurs, the helicity may also decay due to cosmic expansion.

\section{Model with Majorana neutrinos}
\label{sec:majorana}

In this section, we examine the dynamics of the $B-L$ charge in the fermion sector.
We focus 
on the seesaw model\,\cite{Minkowski:1977sc,Yanagida:1979as,Gell-Mann:1979vob}, in which the interaction between the right-handed neutrinos and the standard model particles 
results in the standard model neutrinos behaving as Majorana particles at low energies.
We investigate if our scenario can account for the generation of 
the baryon asymmetry of the universe.

\begin{figure}[t]
    \centering
    \includegraphics[width=0.95\textwidth]{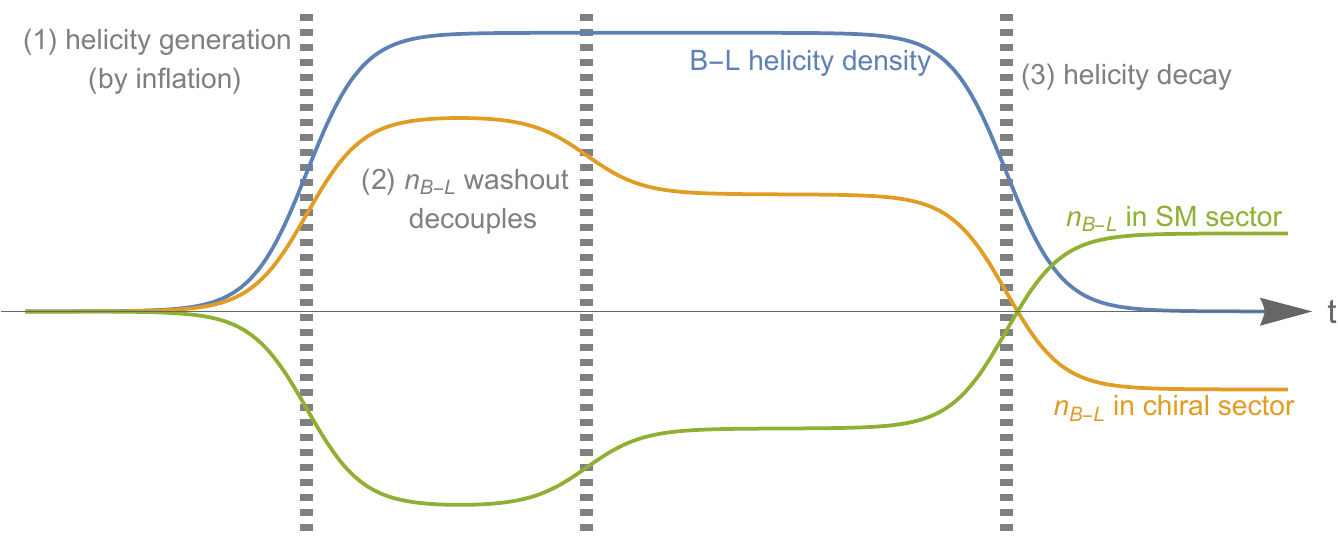}
    \caption{Schematic picture of 
    a successful scenario. First, the $B-L$ helicity (the blue line) is generated by, for example, the axion inflation, together with the $B-L$ charges in the standard model sector (the green line) and the chiral sector (the orange line). Then, the $B-L$ charges are (partially) equilibrated by the interactions between these two sectors to wash out the $B-L$ charge in the fermion sector. Finally, the $B-L$ helicity decays and the $B-L$ charges are generated in both sectors. The final $B-L$ charge in the standard model sector corresponds to the baryon number of the universe today.}
    \label{fig:schematic}
\end{figure}

Let us first recapitulate our leptogenesis scenario and summarize the requirements for the model.
We show a schematic picture of our scenario in Fig.\,\ref{fig:schematic}.
First, we assume the $B-L$ helicity is generated in the early universe. There are several mechanisms to generate the $B-L$ helicity, such as the axion inflation\,\cite{Turner:1987bw,Garretson:1992vt,Anber:2006xt,Barnaby:2010vf,Barnaby:2011vw} and the chiral plasma instability\,\cite{Joyce:1997uy,Akamatsu:2013pjd,Rogachevskii:2017uyc,Schober:2017cdw,Kamada:2018tcs}. As we have discussed in Sec.\,\ref{sec:helicity}, we assume the axion inflation as the generation mechanism in this paper. When the $B-L$ helicity is generated, the $B-L$ charges in the standard model sector are also generated according to the anomaly equation, Eq.\,\eqref{eq:anomaly}.\footnote{In the case of chiral plasma instability~\cite{Joyce:1997uy,Akamatsu:2013pjd,Rogachevskii:2017uyc,Schober:2017cdw,Kamada:2018tcs}, we need to start from a chiral charge for the $B-L$ gauge interaction in the fermion sector. The distribution of $B-L$ charge in the standard model and chiral sector after $B-L$ helicity generation depends on the initial condition.} As long as the $B-L$ symmetry is gauged, regardless of the detail of the model, some chiral sector is added to cancel the 't Hooft anomaly of the $B-L$ symmetry in the standard model sector. The $B-L$ charge is therefore generated in the chiral sector as well. Note that the total $B-L$ charge of the universe is zero; the $B-L$ charges in the standard model sector and the chiral sector are the same amount with the opposite sign. 

As we have discussed in Sec.\,\ref{sec:helicity}, the magnetic helicity of the $B-L$ gauge field decays after the $B-L$ gauge symmetry is Higgsed. The decay generates the $B-L$ charge both in the standard model sector and the chiral sector. If the $B-L$ charges in both sectors are conserved separately from the beginning, the final $B-L$ charges in both sectors are zero. This can be seen from the integrated form of the anomaly equation, Eq.\,\eqref{eq:csconservation}; the initial $B-L$ charges in both sectors and the magnetic helicity are zero and the final helicity is zero, so that the final $B-L$ charges in both sectors are zero.

To generate non-zero $B-L$, we need two requirements.
First, we need to introduce the interactions between the standard model particles and the chiral sector to equilibrate, at least partially, the $B-L$ charges in both sectors, as explained above. However, if this interaction does not decouple even after the decay of the magnetic helicity, the $B-L$ charges generated by the helicity decay in both sectors are also equilibrated, and the final $B-L$ charges in both sectors are zero. Thus, secondly, it is
also required to decouple the interactions before the decay of the magnetic helicity. Then, the decay of the magnetic helicity generates the difference between the $B-L$ charges in the standard model sector and the chiral sector, which remains even today as the baryon number of the universe.

Before discussing the details of the interactions, let us make a few comments on the above summary. Firstly, we have not specified the chiral sector yet. For the seesaw model discussed in this section, the chiral sector is the right-handed neutrinos before the $B-L$ symmetry breaking. However, as the right-handed neutrinos become extremely massive after the symmetry breaking, one may wonder what happens after the decoupling of the right-handed neutrinos. Once the right-handed neutrinos become massive, the $B-L$ symmetry is spontaneously broken and there is a Goldstone boson, $\phi_M$, although the Goldstone boson is eventually eaten by the $B-L$ gauge field. The Goldstone boson has the Wess-Zumino-Witten term and cancels the 't Hooft anomaly\,\cite{Green:1984sg,Faddeev:1986pc,Krasnikov:1985bn,Babelon:1986sv,Harada:1986wb,Preskill:1990fr}. In other words, the Goldstone boson is {\it chiral} and plays the role of the chiral sector. The total $B-L$ charge is always zero even today if we include the unphysical contribution from $\phi_M$, as $B-L$ symmetry is gauged.

Secondly, from the integrated anomaly equation, Eq.\,\eqref{eq:csconservation}, one may wonder if the $B-L$ charge in the standard model sector is always zero given that the initial $B-L$ charge and initial and final helicity is vanishing. Contrary to this naive expectation, the final $B-L$ charge in the standard model sector is not zero.
This is because the interactions equilibrating the $B-L$ charges in the standard model sector and the chiral sector explicitly respect only the total $B-L$ symmetry and break the $B-L$ symmetry in the standard model sector. Schematically, the total derivative of the standard model $B-L$ current is modified as
\begin{align}
    \label{eq:anomaly'}
    \partial_\mu J^\mu_{B-L, \text{SM}} = -\frac{3g_{B-L}^2}{16\pi^2} X_{\mu\nu}\tilde{X}^{\mu\nu} + \mathcal{O}_{\cancel{B-L}}^\text{SM},
\end{align}
where $\mathcal{O}_{\cancel{B-L}}^\text{SM}$ is the operator from the interaction. $\mathcal{O}_{\cancel{B-L}}^\text{SM}$ is, in general, not a total derivative and the sum of the $B-L$ charge in the standard model sector and the helicity is not conserved anymore. In this sense, $\mathcal{O}_{\cancel{B-L}}^\text{SM}$ is the source of the baryon asymmetry of the universe.

With these considerations, we now investigate if the baryon asymmetry of the universe can be generated in the seesaw model through our mechanism. For simplicity, let us first assume that the flavor symmetry is totally broken and the chirality flipping interactions are strong enough and no conserved charge is left in the universe but for the gauge charges. We will shortly discuss the case where the flavor symmetry is not totally broken at the end of this section.

The Lagrangian of the theory is
\begin{align}
    \mathcal{L} &= \mathcal{L}_\text{SM} + \mathcal{L}_N + \mathcal{L}_{N\Phi}+ \mathcal{L}_Y + \mathcal{L}_\Phi + \mathcal{L}_X \label{eq:LUV}\\
    \mathcal{L}_N &\equiv i \bar{N}_i \gamma^\mu\qty(\partial_\mu - i q_N g_{B-L} X_\mu) P_L N_i  \label{eq:LN}\\
    \mathcal{L}_{N\Phi} &\equiv \frac{1}{2}y_{Nij} \Phi N_i^T \mathcal{C} P_L N_j + \text{H.c.} \label{eq:LNPhi}\\
    \mathcal{L}_Y & \equiv y_{Sij}H N_i^T \mathcal{C} L_j + \text{H.c.} \label{eq:LY}\\
    \mathcal{L}_\Phi &\equiv \abs{\qty(\partial_\mu - i q_\Phi g_{B-L} X_\mu)\Phi}^2 - V\qty(|\Phi|^2) \label{eq:LPhi}\\
    \mathcal{L}_X &\equiv -\frac{1}{4}X_{\mu\nu}X^{\mu\nu} \label{eq:LX},
\end{align}
where $N_i$ is $i$-th ($i = 1, 2, 3$) generation of the right-handed neutrino\,\footnote{Readers should not confuse the name, ``right-handed'', with the chirality of the right-handed neutrino. Throughout the paper, we use left-handed Weyl spinors for the chiral fermions.},
$\Phi$ is the Higgs boson of $B-L$, $q_\phi$ is the $B-L$ charge of the field $\phi$, and 
$\mathcal{L}_\text{SM}$ is the standard model Lagrangian. If we ignore the $\theta$ term for $X$, we may assume $y_N$ is diagonal and real, $y_{Nij} = \text{diag}(y_{N1}, y_{N2}, y_{N3})$. For simplicity, we assume all the Yukawa couplings, $y_{Ni}$, are the same order.
After $\Phi$ acquires VEV, $\langle \Phi \rangle = v_{B-L}/\sqrt{2}$, the $B-L$ gauge symmetry is broken and the right-handed neutrinos become massive, leaving the unphysical would-be Goldstone boson, $\phi_M$, as the $B-L$ charge carrier.

The Yukawa interaction between the right-handed neutrinos and the standard model leptons, Eq.\,\eqref{eq:LY}, equilibrate the $B-L$ charges in the standard model sector and the chiral sector. For simplicity, let us first discuss the case where the $B-L$ charges are totally equilibrated.
To estimate the rates, let us take $y_S^2 = y_{S0}^2$ with
\begin{align}
y_{S0}^2 \equiv 2\sqrt{\Delta m_{32}^2} M_{Ni} / v^2 \equiv M_{Ni} / M_\star,
\end{align}
where $\Delta m_{32}$ is the mass difference between the third and the second generation of the neutrinos, $M_{Ni}$ is the zero-temperature mass of the $i$-th right-handed neutrinos which contribute to the equilibrium process the most, and $v$ is the VEV of the Higgs field. Numerically, $M_\star \simeq 6.1 \times 10^{14}\,\text{GeV}$. 
Assuming $y_S \lesssim 1$, the most efficient process to equilibrate the $B-L$ charges is scatterings between the top quark and gauge bosons exchanging the Higgs boson and the lepton. Before the symmetry breaking, the rate of the process is $\Gamma_1 \sim y_{S0}^2 T$, and for $T \ll \frac{M_p}{M_\star}M_{Ni}$, the $B-L$ charges are equilibrated, $\Gamma_1 \gg H(T)$. Therefore, the phase transition temperature of the $B-L$ symmetry breaking, $T_c$, must be much smaller than $\frac{M_p}{M_\star}M_{Ni}$
to meet the first requirement.

\begin{figure}[t]
    \centering
    \includegraphics[width=0.95\textwidth]{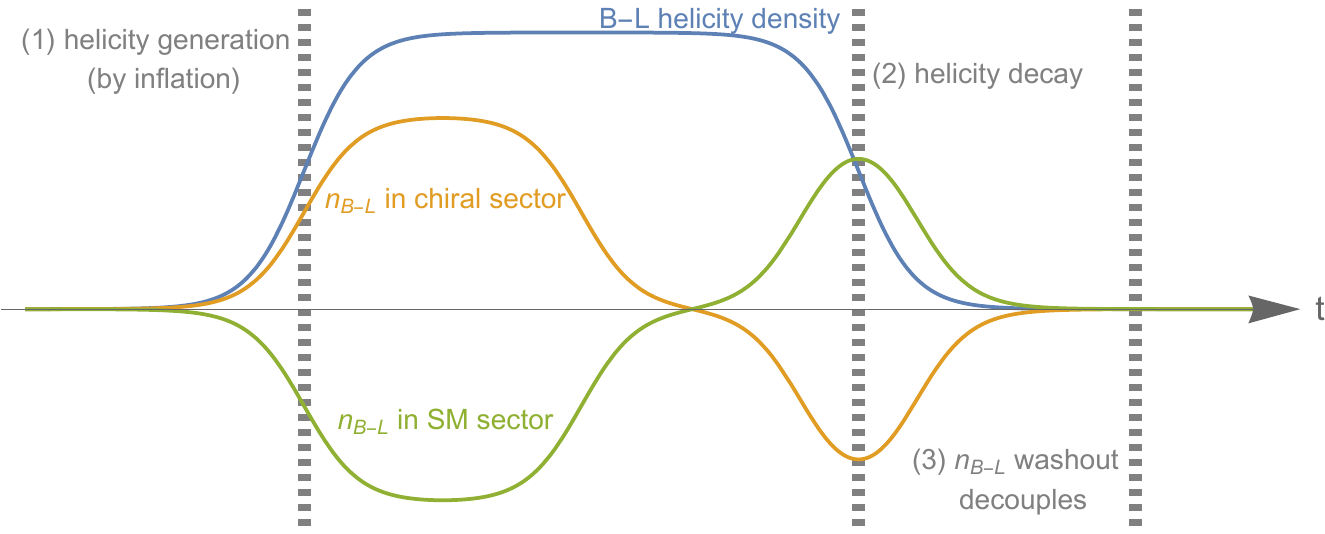}
    \caption{Schematic picture of our scenario with Majorana neutrinos. Compared with Fig.\,\ref{fig:schematic}, the washout process does not decouple before the decay of the magnetic helicity and the final $B-L$ charge in the standard model sector is zero.}
    \label{fig:schematic2}
\end{figure}

On the other hand, the second requirement, decoupling of the interactions before the decay of the magnetic helicity, is almost contradictory to the first requirement. 
As we have discussed in Sec.\,\ref{sec:helicity}, the decay of the magnetic helicity is expected to be quick after the $B-L$ gauge symmetry is broken. Let us denote the temperature $T_{HD}$. Then, the right-handed neutrino is massive and the scattering rate discussed above is suppressed by the Boltzmann suppression factor, $\exp(-M_{Ni}(T_{HD})/T_{HD})$, where $M_{Ni}(T) = y_{Ni}v_{B-L}(T)/\sqrt{2}$. However, as we have discussed in the previous section, $v_{B-L}(T_{HD})/T_{HD}$ is much smaller than unity, and the scattering rate is not suppressed. Therefore, the $B-L$ charges in the standard model sector and the chiral sector are equilibrated even after the decay of the magnetic helicity. We show the schematic picture of our scenario in this case in Fig.\,\ref{fig:schematic2}.

There can be several possibilities to evade this contradiction.
First, we may consider the case where the $B-L$ charges are not totally equilibrated, i.e., the equilibrating interactions are not in the thermal equilibrium. 
Then, the $B-L$ charges in the standard model sector after the phase transition are a tiny fraction of the initial charges with a negative sign and are not washed out. However, the effects of Eq.\,\eqref{eq:LNPhi} become stronger as the temperature decreases compared to the Hubble expansion; if the temperature is smaller than around $y_{S0}^2 M_p = M_{Ni} M_p / M_\star$ well before the right-handed neutrinos decouple, the interaction is in the thermal equilibrium and the final $B-L$ charges in the standard model sector are still washed out.
Second, even when the washout process is in the thermal equilibrium, if it decouples within a few Hubble times after the decay of the magnetic helicity, there could be some asymmetry left in the standard model sector. To study this, we need a more detailed analysis of the helicity decay and the washout process near the phase transition.
Finally, we may consider the case where the phase transition is of the first order. If the system is, for example, overcooled, the washout process in the Higgs phase can be decoupled due to the Boltzmann suppression factor quickly after the phase transition. This, however, requires a more careful analysis of the helicity decay at the first order phase transition, and we leave this for future work.

If we give up the simplest seesaw model and consider more complicated models, we can generate the baryon asymmetry of the universe along our scenario. In the IR limit, the equilibrating interaction is the dimension-5 Weinberg operator, $\mathcal{O}_W = LHLH/\Lambda$, where $\Lambda$ is the cutoff scale. Such a high-dimensional operator is generally stronger in the high-energy scale, but in the simplest seesaw model the operator appears only after the decouple of the right-handed neutrinos, which is the reason why the washout process cannot be decoupled by the time of the helicity decay. Thus, if the dimension-5 operator is generated not by the right-handed neutrinos but by other heavy particles, the washout process can be decoupled earlier and the baryon asymmetry can be generated. For example, we may consider a dimension-6 operator, $\mathcal{O}' = LHLH \Phi^\dagger/\Lambda^2$, which is reduced to the Weinberg operator after the $B-L$ symmetry breaking and assume $y_S$ is much smaller than unity and the neutrino masses are mainly generated by this operator. The $B-L$ equilibration is then achieved by $\mathcal{O}'$ and the washout process can be decoupled by the helicity decay. Such an operator can be generated by introducing new heavy Dirac fermions, for example.

Up to this point, we have ignored the effect of the flavor symmetry of the standard model particles. In the minimal seesaw model, the right-handed neutrino interactions may break the flavor symmetry for the left-handed leptons but the chirality of the right-handed leptons and the flavor symmetry of the quarks are not broken if the corresponding Yukawa couplings are not in the thermal equilibrium\,\cite{Domcke:2020quw,Domcke:2022kfs}. In this case, even though the washout process is strong enough, the chemical potential of the $B-L$ charge and other chiral charges in the standard model sector are not zero before the decay of the magnetic helicity. However, the decay of the magnetic helicity generates the opposite sign of the chemical potentials assuming the washout process is strong enough. Therefore, the final $B-L$ charge in the standard model sector is still zero.

\section{Model with Dirac neutrinos}
\label{sec:dirac}
In this section, we focus on the model with Dirac neutrinos to overcome the problem in the minimal seesaw model. We will show that in this case
the baryon asymmetry of the universe can be generated in the model through our mechanism. The minimal Lagrangian of the model is
\begin{align}
    \mathcal{L} &= \mathcal{L}_\text{SM} + \mathcal{L}_N + \mathcal{L}_Y + \mathcal{L}_\Phi + \mathcal{L}_X.
\end{align}
The neutrino mass is generated by the Yukawa interaction between the right-handed neutrinos and the standard model leptons, Eq.\,\eqref{eq:LY} and the neutrino is now Dirac. The masses are
\begin{align}
    m_{\nu i} = y_{Si} \frac{v}{\sqrt{2}},
\end{align}
where we diagonalize the Yukawa coupling $y_{Sij} = U_{ik}^{N} \text{diag}\qty(y_{S1}, y_{S2}, y_{S3})_{kl} U_{lj}^{\nu}$ by the unitary matrices $U^N$ and $U^\nu$ so that $y_{S3} > y_{S2} > y_{S1} \geq 0$. To reproduce the observed neutrino masses, we require
\begin{align}
    y_{S3} \simeq \frac{\sqrt{2}\Delta m_{32}^2}{v} = 2.8 \times 10^{-13}.
\end{align}
Hence, the Yukawa interaction is decoupled in the universe.

As we have discussed in the previous section, we need two requirements to generate the baryon asymmetry of the universe. The first requirement is the washout of the $B-L$ charges in the standard model sector and the chiral sector. The second requirement is the decoupling of the interactions before the decay of the magnetic helicity. As the Yukawa interaction is decoupled, the first requirement is not satisfied, because the Lagrangian respects the global $B-L$ symmetry. 
However, the Yukawa interaction is not the only possible interaction between the right-handed neutrinos and the standard model leptons. For example, we can 
introduce higher-dimensional operators between the right-handed neutrinos and the standard model leptons. An example of the relevant operators consistent with the symmetries is a dimension-8 operator,
\begin{align}
    \mathcal{L}_\text{dim-8} = \frac{1}{\Lambda^4} \qty(HN_i^T \mathcal{C} L_j)^2 + \text{H.c.},
\end{align}
where $\Lambda$ is the cutoff scale. In general, the cutoff scale is expected to be the Planck scale, $\Lambda \sim M_p$. If we introduce additional heavy fields, the cutoff scale can be much smaller than the Planck scale. 
Another possibility, as seen in some models called the Dirac seesaw models (e.g., 
Ref.\,\cite{Roncadelli:1983ty,Ma:2014qra}), involves the introduction of new heavy particles 
that equilibrate the $B-L$ charges in the standard model sector and chiral sector. As these models introduce additional charged particles and the dynamics of the $B-L$ gauge field change, for simplicity, we focus on the simplest model with higher-dimensional operators. With higher-dimensional operators, the $B-L$ charges in the standard model sector and the chiral sector are at least partially equilibrated. Let $c$ denote the fraction of the $B-L$ charge washed out; $c=1$ if the $B-L$ charges are totally equilibrated.

\begin{figure}[tb]
    \begin{minipage}{0.45\hsize}
        \centering
        \includegraphics[width=0.95\textwidth]{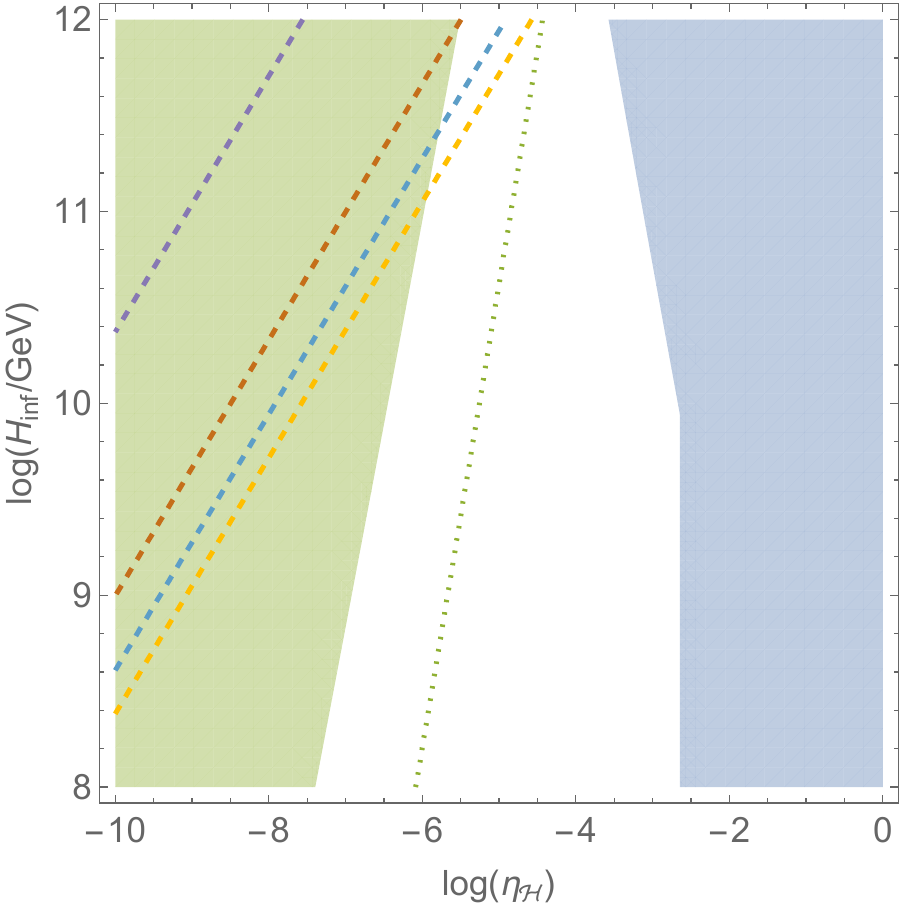}
    \end{minipage}
    \begin{minipage}{0.45\hsize}
        \centering
        \includegraphics[width=0.95\textwidth]{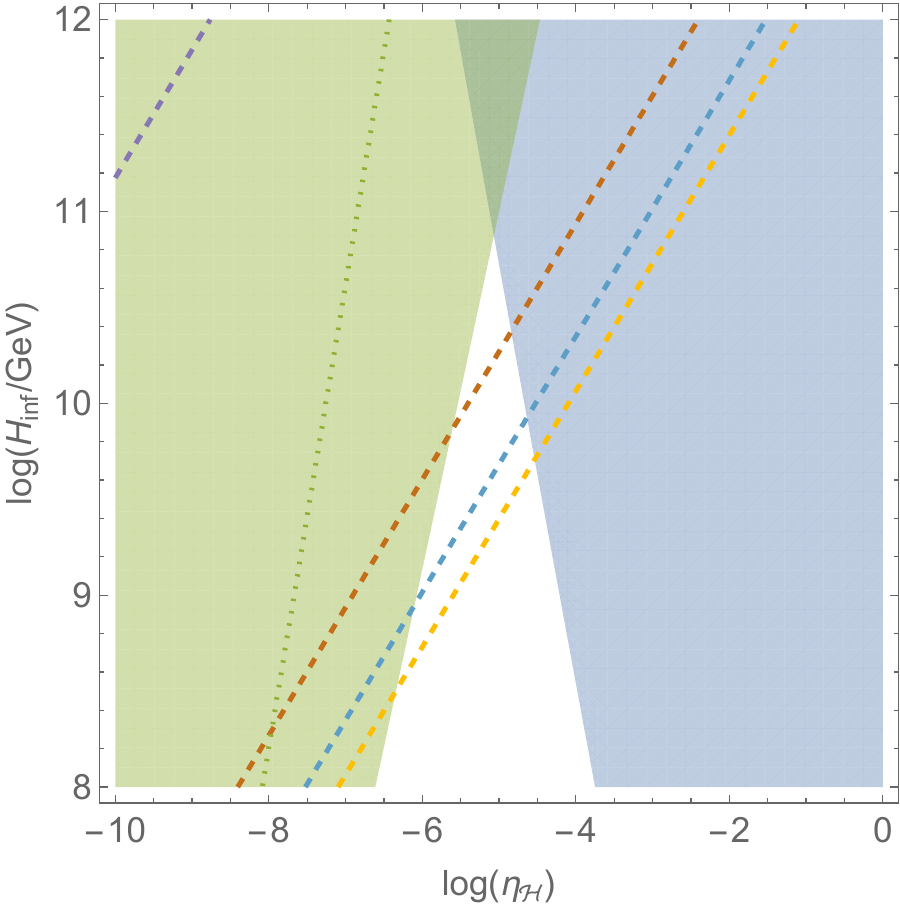}
    \end{minipage}
    \caption{The excluded region of the parameter space in terms of the $B-L$ asymmetry and the Hubble constant at the end of the inflation. The left panel is for $g_{B-L} = 0.1$ and the right panel is for $g_{B-L} = 0.01$. The blue region is the excluded region by the energy density of the magnetic field and the generated Fermions. The green region is the excluded region by the requirement of the magnetic Reynolds number, $R_m < 1$. We also indicate the contour with $R_e = 1$ by the green dotted line. The region left to the green dotted line is the region where the kinetic Reynolds number is smaller than unity.
    The purple, brown, light blue, and yellow dashed lines are the contours of $\xi = 5, 10, 15, 20$, respectively.}
    \label{fig:result}
\end{figure}

In Fig.\,\ref{fig:result}, we show the excluded region of the parameter space in terms of the $B-L$ asymmetry and the Hubble constant at the end of the inflation, assuming that the magnetic helicity is generated during the axion inflation as discussed in Sec.\,\ref{sec:helicity}. 
The blue region is the excluded region where the energy density of the magnetic field and the generated asymmetric fermion are larger than $1\,\%$ of the total energy density of the universe at the end of the inflation. The green region is the excluded region by the requirement of the magnetic Reynolds number, $R_m \lesssim 1$. We also show the contour with $R_e = 1$ by the green dotted line as a reference. The region left to the green dotted line is the region where the kinetic Reynolds number is smaller than unity.

From the figure, we can see that we need a larger $B-L$ helicity for the magnetic field not to dissipate
for a conservative estimate of the magnetic Reynolds number adopted in Sec.~\ref{sec:helicity}. This requires that the fraction of the $B-L$ charge washed out, $c$ is much smaller than unity, as $c \simeq \frac{79}{28} \eta_0 / \eta_{\cal H}$ to explain the baryon asymmetry of the universe, assuming that the anomalous charge cancellation does not occur. Alternatively, we can dilute the baryon asymmetry by the entropy production after the decay of the magnetic helicity.
Another possibility is to consider the case where the helicity is decaying by the dissipation or the anomalous charge cancellation, but the final $B-L$ charge is determined by the balance between the decay of the helicity and the washout process. 
Once more, since the condition adopted here for the magnetic field not to dissipate is a rough 
estimate, the possibility that the magnetic field survives at smaller $\eta_H$ is not completely ruled out~\cite{Domcke:2022kfs}.  Determining the parameter space for the 
successful baryogenesis scenario
requires  
more detailed analyses of the helicity evolution and decay by MHD as well as fermion production, which are beyond the scope of this paper. 
To conclude this section, we note that the generation of the baryon asymmetry of the universe in the model with Dirac neutrinos can be seen as an implementation of Dirac leptogenesis\,\cite{Dick:1999je,Murayama:2002je}. In Dirac leptogenesis, the decay of a heavy particle produces a $B-L$ asymmetry in both the standard model sector and the right-handed neutrino sector. In our scenario, this asymmetry is generated through the decay of the magnetic helicity of the $B-L$ gauge field.

\section{Discussion and Conclusion}
\label{sec:conc}

In this paper, we have proposed a new scenario to generate the baryon asymmetry of the universe. We have discussed the decay of the magnetic helicity of the $B-L$ gauge field, which is generated, e.g., during axion inflation, after the $B-L$ gauge symmetry is Higgsed. 
The net $B-L$ asymmetry in the standard model sector is the combination of the asymmetry generated during inflation, which is subject to the washout effect, and the asymmetry generated 
by the $B-L$ helicity decay.
For both the seesaw model with Majorana neutrinos and the model with Dirac neutrinos, we have investigated if the baryon asymmetry of the universe can be generated in the model through our mechanism. We have shown that the observed baryon asymmetry can be generated in the model with Dirac neutrinos if the effect of the washout process is decoupled before the decay of the magnetic helicity.
For the model with Majorana neutrinos, we have shown that the baryon asymmetry of the universe cannot be generated in the minimal seesaw model with the second-order phase transition, although the baryon asymmetry can be generated in more complicated models.

While our discussion on the helicity evolution is based on order-of-magnitude estimates from MHD, the fundamental principles of our scenario remain robust. These principles include the conservation of $B-L$ magnetic helicity, the equilibration and subsequent decoupling of $B-L$ charges between the standard model sector and the chiral sector, and the decay of magnetic helicity following the $B-L$ phase transition. However, to precisely quantify the baryon asymmetry of the universe, detailed analytical formulation and numerical simulations of helicity decay, particularly near the phase transition, are essential.

We have focused on the axion inflation as the generation mechanism of the magnetic helicity. However, there can be other mechanisms to generate the magnetic helicity, such as the chiral plasma instability\,\cite{Joyce:1997uy,Akamatsu:2013pjd,Rogachevskii:2017uyc,Schober:2017cdw,Kamada:2018tcs} and the kinetic misalignment\,\cite{Co:2022kul}. It is interesting to investigate if the baryon asymmetry of the universe can be generated in these scenarios as well.

Finally, in this paper, we have
ignored the effect of the $\text{U}(1)_Y$ gauge field and and its dynamics. For example, the kinetic mixing between the $B-L$ gauge field and the $\text{U}(1)_Y$ can generate the helicity of the $\text{U}(1)_Y$ gauge field~\cite{Kamada:2018kyi}. Also, if the generated $B-L$ charge is large and the temperature is low enough, the chiral plasma instability may generate the helicity of the $\text{U}(1)_Y$ gauge field. It is interesting to investigate the coupled dynamics of the $B-L$ and $\text{U}(1)_Y$ gauge fields and the baryon asymmetry of the universe in this case, although we leave this for future work.

\acknowledgments
The work of H.F. was supported by JSPS KAKENHI Grant No.\ 24K17042.
The work of K.K. was supported by the National Natural Science Foundation of China (NSFC) under Grant No.~12347103 and JSPS KAKENHI Grant-in-Aid for Challenging Research (Exploratory) JP23K17687.
The work of T.S. was supported by the JSPS fellowship Grant No.\ 23KJ0678.

While finishing this work, we have noticed that a related idea is discussed in Ref.\,\cite{Chao:2024fip}.

\bibliography{ref}

\end{document}